# Interspecies collision-induced losses in a dual species $^7$Li-$^{85}$Rb magneto-optical trap

Sourav Dutta[1,*], Adeel Altaf[1], John Lorenz[1], D. S. Elliott[1,2,‡], and Yong P. Chen[1,2,†]

[1] Department of Physics, Purdue University, West Lafayette, IN 47907, USA
[2] School of Electrical and Computer Engineering, Purdue University, West Lafayette, IN 47907, USA

In this article, we report the measurement of collision-induced loss rate coefficients $\beta_{Li,Rb}$ and $\beta_{Rb,Li}$, and also discuss means to significantly suppress such collision induced losses. We first describe our dual-species magneto-optical trap (MOT) that allows us to simultaneously trap $\geq 5\times10^8$ $^7$Li atoms loaded from a Zeeman slower and $\geq 2\times10^8$ $^{85}$Rb atoms loaded from a dispenser. We observe strong interspecies collision-induced losses in the MOTs which dramatically reduce the maximum atom number achievable in the MOTs. We measure the trap loss rate coefficients $\beta_{Li,Rb}$ and $\beta_{Rb,Li}$, and, from a study of their dependence on the MOT parameters, determine the cause for the losses observed. Our results provide valuable insights into ultracold collisions between $^7$Li and $^{85}$Rb, guide our efforts to suppress collision induced losses, and also pave the way for the production of ultracold $^7$Li$^{85}$Rb molecules.

## 1. INTRODUCTION

Simultaneous cooling and trapping of two or more species of alkali atoms has attracted great interest in recent years [1-22]. Systems with two species have found application in sympathetic cooling [22-24] and have provided a wealth of information on ultracold collisions [6-19]. Such two species systems also form the starting point for most experiments designed to create ultracold ground state heteronuclear polar molecules [1-5]. Ground state heteronuclear molecules have recently attracted enormous attention, for example due to their electric dipole moment as a possible basis for quantum computing protocols [25, 26]. In addition, polar molecules also provide a good system for precision measurements [27] and for studying ultracold chemistry [28-30], quantum phase transitions and quantum simulations [31, 32]. The value of the electric dipole moment of heteronuclear molecules is generally the highest and decoherence is lowest when the molecules are in their ro-vibronic ground state. This has led to an increased interest in creating ultracold heteronuclear molecules in their ro-vibronic ground state starting from two co-trapped species of atoms with the primary methods being photo-association (PA) [3, 4, 33] and magneto-association (MA) followed by Stimulated Raman Adiabatic Passage (STIRAP) [1, 2].

Our interest in cooling and trapping $^7$Li and $^{85}$Rb atoms stems from the relatively high value of the electric dipole moment of 4.1 Debye for LiRb molecules in the ro-vibronic ground state [34]. The production of LiRb molecules in their ro-vibrational ground state requires the knowledge of the potential energy curves and such information has recently started becoming available through high resolution spectroscopic measurements of LiRb molecules in hot gaseous phase [35-37]. Recently, two groups have also reported observation of Feshbach resonances in the $^6$Li-$^{85}$Rb, $^6$Li-$^{87}$Rb and $^7$Li-$^{87}$Rb systems [38-40].

Studies of collisions between ultra-cold atoms are of both fundamental and practical importance. On the fundamental side, they have provided important information regarding molecular potential energy curves [5, 6] and identification of Feshbach resonances [38-40]. Isotopic differences for various species have also been revealing, in that they can shed light on the role of hyperfine-structure changing collisions or subtle energy differences between high-lying vibrational states. Such effects have been observed in potassium [18] ($^{39}$K vs. $^{41}$K), rubidium [19] ($^{85}$Rb vs. $^{87}$Rb), and, in heteronuclear systems, RbCs [14] ($^{85}$Rb$^{133}$Cs vs. $^{87}$Rb$^{133}$Cs) and LiRb [10] ($^6$Li$^{85}$Rb vs. $^6$Li $^{87}$Rb). As a practical aspect, one must understand and quantify collision effects in a MOT system in order to optimize the number of trapped atoms, temperature, and the atomic density of the MOT.

In this article, we report a dual-species magneto-optical trap (MOT) for simultaneous cooling and trapping of $^7$Li and $^{85}$Rb, aimed at creating ultracold polar $^7$Li$^{85}$Rb molecules. We describe the dual-species MOT apparatus, which allows us to simultaneously trap $\geq 5\times10^8$ $^7$Li atoms loaded from a Zeeman slower and $\geq 2\times10^8$ $^{85}$Rb atoms loaded from a dispenser. We have observed interspecies collision-induced losses in the MOTs, measured the trap loss rate coefficients $\beta_{^7Li^{85}Rb}$ and $\beta_{^{85}Rb^7Li}$, and studied their dependence on the MOT parameters. Our results show that the primary loss mechanism in the Li-Rb system is due to collisions between excited state Rb atoms and ground state Li atoms. In this regard, the Li-Rb system is similar to that of Li-Cs [9]. In order to explore isotopic effects, we compare our results to those of Ref. [10] for the $^6$Li-$^{85}$Rb system. Comparison of the relative loss rates for Rb-induced Li losses vs. Li-induced Rb losses suggests an enhancement of LiRb molecular association by the Rb trapping beams.

In the following section, we describe the dual species MOT system in which we carry out these measurements of collision-induced losses. In Section 3, we describe the collision measurements, and discuss our analysis and interpretation of the collision-induced loss rates. We then discuss means to strongly suppress such collision-induced losses, followed by the conclusion.

## 2. EXPERIMENTAL SETUP

### 2.1 Light sources for the MOT

A schematic representation of the laser system is shown in Fig. 1. To drive transitions between the $5s\ ^2S_{1/2}$ and $5p\ ^2P_{3/2}$ states, the $^{85}$Rb MOT requires two lasers with wavelength near 780 nm: a cooling laser and a repumping laser, differing in frequency by the ground state hyperfine splitting (~ 3.036 GHz). The cooling laser is a commercial high power (~ 1W) external cavity diode laser (ECDL) from Sacher Lasertechnik. The repumping laser is a homebuilt ECDL in a Littrow configuration with an output power of around 50 mW [41, 42]. Frequency stabilization of both lasers is obtained by locking the laser frequency using the standard saturated absorption spectroscopy technique. The cooling laser is locked to the $F = 3 \to F' =$ 2-4 crossover resonance. The frequency is then up-shifted by 68 MHz using an acousto-optic modulator (AOM) in a single pass configuration, which makes the cooling laser frequency detuned by $\delta_{Rb} = -24$ MHz from the $F = 3 \to F' = 4$ cycling transition. The homebuilt repumping laser is locked to $F = 2 \to F' =$ 1-2 crossover resonance. The frequency is then up-shifted by 78 MHz using another AOM in a single pass configuration, which makes the repumping laser resonant with the $F = 2 \to F' = 3$ transition. The repumping light is combined with the cooling light on a polarizing beam splitter (PBS). Both beams are then sent to a dichroic mirror where they are combined with the 671nm light for the $^7$Li MOT.

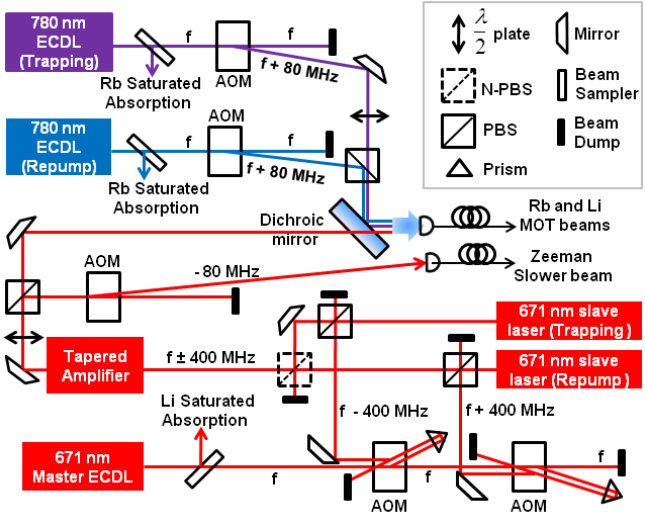

FIG. 1. (Color online) Schematics of the laser system for our $^7$Li-$^{85}$Rb dual species magneto-optical trap (MOT). For the AOMs in double pass configuration, right angled prisms are used to vertically displace the beams. The beams coming out of the two slave lasers are combined on an N-PBS instead of a PBS because the tapered amplifier can amplify light of only one polarization. Note that the cooling and repumping beams for both $^7$Li and $^{85}$Rb MOTs are combined and sent to the experiment through the same fiber.

To drive transitions between the $2s\ ^2S_{1/2}$ and $2p\ ^2P_{3/2}$ states, the $^7$Li MOT also requires light at two frequencies, for the cooling and repumping transitions, separated by the $^7$Li ground state hyperfine splitting of ~803.5 MHz. The Lithium laser system is based on a master-slave injection scheme. We use a commercial ECDL (Toptica DLPro) as our master laser with ~20mW output power at 670.96 nm. The master laser is locked to the $F = 1$-$2 \to F'$ crossover resonance in the saturated absorption spectra of $^7$Li (note that the hyperfine levels in the $2p\ ^2P_{3/2}$ state of $^7$Li are not well resolved). To generate the frequency at the cooling (repumping) $F = 2 \to F' = 3$ ($F = 1 \to F' = 2$) transition, a part of the light from the master laser is down-shifted (up-shifted) in frequency by ~ 400 MHz using a 200 MHz AOM in a double pass configuration. The down-shifted and up-shifted beams are used to injection lock two free running laser diodes, each producing ~20 mW of light at the frequency of the respective transitions. We control the detuning of the cooling and repumping beams using their respective AOMs. The outputs from the two injection locked lasers are combined on a non-polarizing beam splitter (N-PBS). The ratio of power between the cooling and repumping light is controlled by suitably placed half-wave plates and polarizing beam splitters (PBS) before the beams are combined on the N-PBS. The combined light, containing both cooling and repumping frequencies, is injected into a commercial tapered amplifier (Toptica BoosTA), which produces up to 270 mW of light. The spectral content of the tapered amplifier output is checked with a scanning Fabry-Perot interferometer with free spectral range of 2 GHz. We adjust the injected power such that the height of the transmission peak at the cooling frequency is twice that at the repumping frequency. This determines the ratio, 2:1, between the powers in the cooling and repumping frequencies with the ratio fixed for the experiments reported here. The light from the tapered amplifier is divided into two parts. The major part (~180 mW) is sent to a dichroic mirror where it is combined with the 780 nm light. The beams for the two MOTs are coupled into the same polarization maintaining optical fiber, which greatly simplifies the optical set up near the vacuum chamber. We typically get around 50% coupling efficiency for both the 671 nm and 780 nm beams, yielding up to 90 mW of light at 671 nm and up to 300 mW of light at 780 nm. The other part of the light from the tapered amplifier, comprised of both cooling and repumping frequencies, is down-shifted in frequency by 80 MHz using an AOM to provide the light for the Zeeman slower. After coupling into a polarization maintaining optical fiber, the maximum power available for the Zeeman slower beams is ~20 mW. Note that this total power in the Zeeman slower beam is also distributed between two frequencies. In the rest of the article we often drop the superscripts and denote $^7$Li with Li and $^{85}$Rb with Rb.

### 2.2 Vacuum chamber

We show a schematic diagram of the entire vacuum chamber in Fig. 2. The vacuum chamber consists of three main sections: the Lithium oven, the Zeeman slower for Li atoms and the ultra high vacuum (UHV) experimental chamber.

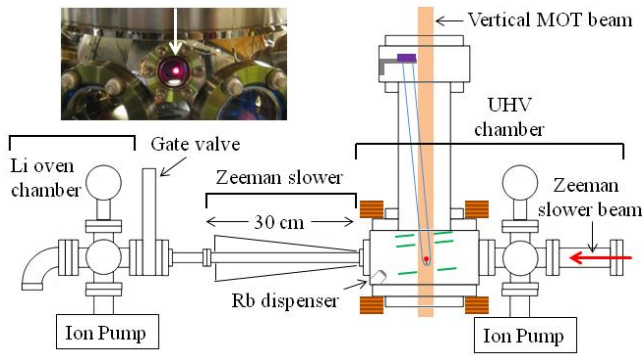

FIG. 2. (Color online) A schematic representation of the vacuum chamber. The red dot at the center of the UHV chamber represents the MOT. The green lines denote the electric field plates which, in future experiments, will accelerate the ions produced during REMPI towards the MCP, denoted in purple. The trajectory of ions is indicated by the blue lines. The MOT coils are denoted in orange. Inset: Photograph of the $^7$Li MOT (indicated by the arrow).

The lithium (Li) oven chamber (Fig. 3(A)) produces a collimated beam of Li atoms travelling toward the UHV chamber via the Zeeman slower section. It consists of a Li oven containing approximately 10g of lithium (natural abundance, ~ 92% $^7$Li) which is heated to ~ 400°C resulting in a lithium vapor pressure of $10^{-4}$ Torr. The hot vapor escapes the oven through a nozzle (diameter = 8 mm), which is kept at a slightly higher temperature of ~ 415°C to avoid condensation of Li. As shown in Fig. 3(B), the nozzle consists of a stack of approximately 60 hypodermic needles, each with an inner diameter of 0.8 mm and length of 10 mm. The hypodermic needles improve the collimation of the atomic beam by reducing the emission angle to ~ 4.5°. The atomic beam is further collimated by an aperture of 8 mm diameter placed approximately 80 mm downstream from the nozzle. As shown in Fig. 3(C), the aperture is formed by two holes on a hollow cylinder mounted on a vacuum rotation feedthrough. In addition to its role in improving the beam collimation (when the aperture faces the atomic beam), it serves as a beam shutter, completely blocking the atomic beam when the aperture is perpendicular to the atomic beam. In addition, it can also control (reduce) the atomic flux by appropriate rotation of the hollow cylinder. The collimated Li beam then enters the Zeeman slower section after passing through a gate valve. The oven chamber is pumped by a Varian Starcell Ion pump with a pumping speed of 40 l/s. To protect the gate valve from direct contact with lithium atoms, the copper gasket forming the vacuum seal is slightly unconventional. Instead of the standard copper gaskets, a blank copper gasket is modified to include a small through hole of 8 mm diameter at the center. This hole also provides additional collimation to the lithium beam and reduces the conductance between the oven chamber and the Zeeman slower section. The thermal lithium atoms emanating from the oven chamber are slowed by the Zeeman slower [43]. We show the magnetic field of the Zeeman slower along its axis in Fig. 3 (D). The net magnetic field is the sum of the magnetic field produced by the 8-section Zeeman slower solenoid with variable number of turns and by the MOT coils

[44]. The Zeeman slower is in a decreasing field configuration with maximum magnetic field near the Li oven and decreasing to zero near the MOT. The 12" long tube of the Zeeman slower has an inner diameter of only 0.75" resulting in a low conductance between the oven chamber and the UHV experimental chamber. We estimate that the low conductance helps maintain the UHV chamber at a pressure 50 times lower than the oven chamber.

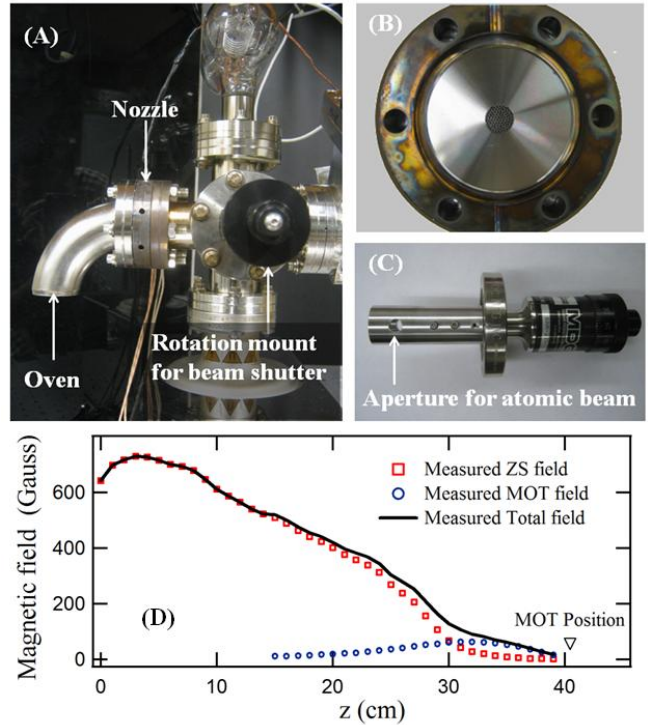

FIG. 3. (Color online) (A) Photograph of the lithium oven chamber showing the oven, the nozzle and the rotation mount on which the hollow cylindrical beam shutter is mounted (see text for details). (B) Photograph of the nozzle before integration into the vacuum chamber. The stacked hypodermic needles in the central aperture are also visible. (C) Photograph of the atomic beam shutter mounted on a rotation mount. The atom beam is blocked or transmitted depending on the orientation of the aperture. (D) Magnetic field profile of the Zeeman Slower (ZS) solenoid (squares), MOT coils (circles) and their combined magnetic field (solid line).

The UHV experimental chamber (Fig. 2) is the heart of the experiment and is designed to produce and detect ultracold atoms and molecules. The pressure in the UHV experimental chamber is less than the lowest pressure, $4 \times 10^{-10}$ Torr, measurable by the ion gauge. As shown in Fig. 2, the UHV chamber consists of an extended 8" spherical octagon (Kimball Physics MCF800-ExtOct-G2C8A16), a 10.6" long CF 6" nipple and a 6" spherical octagon (Kimball Physics MCF600-SphOct-F2C8) on the top. The extended 8" spherical octagon has two CF 8" viewports, eight CF 2.75" viewports and sixteen CF 1.33" viewports. This allows for excellent optical access. The two CF 8" viewports are used for the vertical MOT beams. Four of the CF 2.75" viewports are used for the horizontal MOT beams, two are used for fluorescence imaging, one connects to the Zeeman slower section and the last connects to a six-way cross. The arms of the six-way cross are connected to a Varian Starcell Ion pump, an ion gauge and a sapphire window through which the laser beam for the Zeeman slower enters. All other viewports are standard Kodial glass viewports (Kurt J. Lesker) which were broadband anti-reflection coated for the 650-1100 nm region by Abrisa. Ten out of the sixteen CF 1.33" viewports have optical viewports, two have electrical feedthroughs while the rest are blanked off. The Rb MOT is loaded from a Rb dispenser (SAES Getters) located approximately 6 cm from the MOT. The Rb dispenser is typically operated by running a current of 3.3 A. The pressure of the UHV chamber increases to ~ $3\times10^{-9}$ Torr when the Rb dispenser is in operation.

The apparatus is designed for experiments to produce ultracold LiRb molecules in their ro-vibronic ground state. The LiRb molecules formed in our experiments will be detected using Resonance Enhanced Multi Photon Ionization (REMPI). Details of REMPI are outside the scope of this article but similar schemes have been discussed elsewhere [45]. A Time-of-Flight (TOF) Mass Spectrometer [46] installed inside the UHV chamber will be used to detect and image the ions formed during REMPI. A relatively new design feature of our apparatus is the ability to detect the orientation of the LiRb molecules. This will be achieved using the technique of Velocity Mass Imaging (VMI) [47]. We integrate the TOF MS and VMI into one compact set-up. The details of the molecules detection set-up will be discussed in a future report.

### 2.3 Two-species Magneto-optical trap

As discussed above, the light for the $^7$Li and $^{85}$Rb MOTs is coupled into a single optical fiber and brought to the table on which the experiments are performed. This ensures good overlap between the 671 nm and 780 nm beams in addition to a pure Gaussian beam profile. The MOTs are formed by three retro-reflected pairs of mutually perpendicular laser beams intersecting at the center of the UHV chamber. The beams have a $1/e^2$ diameter of ~ 22 mm. The appropriate circular polarizations [19] of the laser beams are obtained using achromatic quarter wave-plates. The MOTs are operated at an axial (i.e. vertical) magnetic field gradient of ~11 Gauss/cm provided by a pair of current carrying coils. It may be noted here that these coils are not exactly in an anti-Helmholtz configuration, resulting in a horizontal magnetic field gradient less than that expected for anti-Helmholtz coils. We fine tune the overlap of the two MOTs by minor adjustments of the quarter wave-plates. The numbers of trapped atoms change only slightly on adjustment of the quarter wave-plates. When the maximum available laser powers (100 mW (30 mW) in each of the six beams for Rb (Li) MOT) are used, we are able to trap $\geq 2\times10^8$ Rb atoms and $\geq 5\times10^8$ Li atoms with typical densities of ~ $3\times10^9$ cm$^{-3}$ and ~ $2\times10^{10}$ cm$^{-3}$ respectively. Upon reduction of the cooling beam power, we can control (reduce) the MOT size and atom number, facilitating the measurements of the collision rates described in the following section.

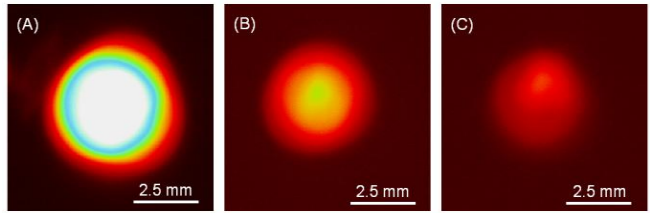

FIG. 4. (Color online) False color fluorescence images of the MOTs taken by one of the CCD cameras. (A) Rb MOT in absence of Li MOT, (B) Li MOT in the absence of the Rb MOT, and (C) Li MOT in the presence of the Rb MOT (a filter is used to block most of the Rb fluorescence). The reduction in the number of trapped Li atoms due to the presence of the Rb MOT is clearly visible.

The numbers of atoms in the MOTs are monitored by fluorescence detection [48]. The fluorescence from both MOTs is collected using a pair of lenses. The Li and Rb fluorescence are separated using a dichroic mirror and detected with two separate large-area photodiodes. Around 3.5% (1.5%) of the Li (Rb) fluorescence leaks into the Rb (Li) detection channel. The spurious signal is subtracted from the recorded signal resulting in negligible cross-talk between the Li and Rb detection channels. In addition to the photodiodes, two CCD cameras are used to record the images of both MOTs from two orthogonal directions. The CCD camera images are used to measure the sizes of the two MOTs (used to infer the atom densities) and to monitor their spatial overlap (see Fig. 4). Typical sizes of the MOTs are ~3 mm, and the Li and Rb MOTs are very well overlapped.

## 3. RESULTS AND ANALYSIS

### 3.1 Measurement and analysis of loss rates

Collisions between Li and Rb atoms in the dual-species MOT lead to loss of atoms from the MOT. As a result, the steady state atom number in one MOT is reduced when the MOT of the other species is present. Figure 5 shows an example of the Li and Rb MOT fluorescence signals corresponding to the following loading sequence. Initially the Li light (both cooling & repump) and the Rb repumping light are blocked and none of the MOTs are loaded (the Rb cooling light is always on). At $t$ = 10 s, the Li light is unblocked allowing the Li MOT to load. After the Li MOT reaches its steady state, the Rb repumping light is unblocked at $t$ = 40 s allowing the Rb MOT to load in presence of the Li MOT. The number of atoms in the Li MOT is now reduced in the presence of the Rb MOT and reaches a new steady state. At $t$ = 70 s, the Li light is blocked to remove the Li MOT resulting in an increase of the atom number in the Rb MOT. At $t$ = 90 s, both the Li and Rb beams are blocked. The loading sequence is then reversed.

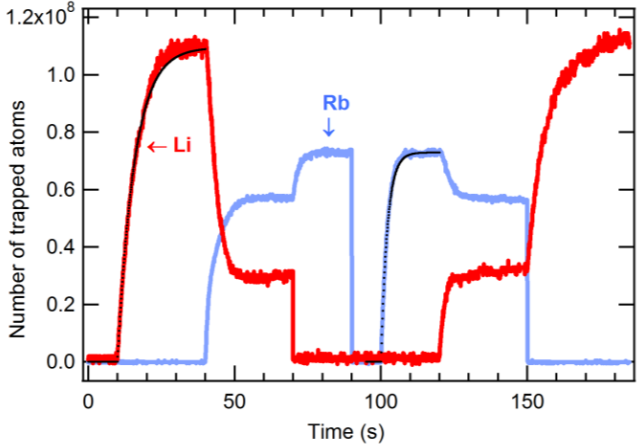

FIG. 5. (Color online) Li and Rb MOT fluorescence signal. The loading sequence is described in the text. The black dotted lines are the fits to equation (5) for single species MOT loading.

The loading of a MOT of species $A$ in the presence of species $B$ can be modeled by the rate equation [7]:

$$\frac{dN_A}{dt} = L_A - \gamma_A N_A - \beta_A \int n_A^2 d^3r - \beta_{A,B}\int n_A n_B d^3r \quad (1)$$

where $N_A$ is the number of atoms in the species $A$ MOT, $n_A$ and $n_B$ are densities of MOTs of species $A$ and $B$ respectively and $L_A$ is the loading rate for species $A$. $\gamma_A$ is the 1-body loss rate coefficient accounting for the losses due to collisions with the background gases, $\beta_A$ is the 2-body loss rate coefficient accounting for the losses of species $A$ due to collisions between atoms of species $A$, and $\beta_{A,B}$ is the 2-body loss rate coefficient accounting for the losses of species $A$ due to collisions with species $B$. The order of indices in $\beta_{A,B}$ is relevant with the first index standing for the species being lost due to the presence of the species indicated by the second index. The analysis of loss rates using the above equation is simplified by the following two conditions which are maintained in our experiments: (i) the MOTs operate in the constant density regime, generally true for MOTs with $10^5$ or more atoms [49,50], where the density of the MOT remains approximately constant during the loading of the MOT while the volume increases, allowing the simplification: $\beta_A \int n_A^2 d^3r = \beta_A n_A N_A$, and (ii) one of the MOTs (say $A$) is smaller than the other MOT (say $B$) allowing the simplification: $\beta_{A,B}\int n_A n_B d^3r = \beta_{A,B} n_B N_A$.

To obtain the value of $\beta_{Li,Rb}$, a small Li MOT is loaded in the presence of a bigger Rb MOT. A smaller Li MOT is obtained either by turning the Zeeman slower magnetic field off or by reducing the power of the Li cooling laser or both. Under these conditions, Eq. (1) can be written as:

$$\frac{dN_{Li}}{dt} = L_{Li} - (\gamma_{Li} + \beta_{Li} n_{Li}) N_{Li} - \beta_{Li,Rb} n_{Rb} N_{Li} \quad (2)$$

To obtain the value of $\beta_{Rb,Li}$, a small Rb MOT is loaded in the presence of a bigger, Zeeman slower-loaded Li MOT.

A smaller Rb MOT is loaded by reducing the power of the Rb cooling laser. Under this condition, Eq. (1) can be written as:

$$\frac{dN_{Rb}}{dt} = L_{Rb} - (\gamma_{Rb} + \beta_{Rb} n_{Rb}) N_{Rb} - \beta_{Rb,Li} n_{Li} N_{Rb} \quad (3)$$

These equations, (2) and (3), can also be used to describe the loading of a single species MOT by setting the last term to zero, leading to equations of the type:

$$\frac{dN_A}{dt} = L_A - \kappa_A N_A \quad (4)$$

where $\kappa_A = (\gamma_A + \beta_A n_A)$. The solution to this equation is:

$$N_A(t) = N_A^\infty (1 - e^{-\kappa_A t}) \quad (5)$$

where $N_A^\infty = L_A/\kappa_A$ is the number of atoms in the steady state MOT of species $A$ in the absence of MOT of species $B$. The values of $\kappa_{Li}$, $\kappa_{Rb}$, $L_{Li}$ and $L_{Rb}$ are obtained from a fit of Eq. (5) to the experimental loading data for single species MOT. The values depend on the detuning of the respective MOT lasers. For $\delta_{Li}$ = -9 – -39 MHz, typical values are: $\kappa_{Li}$ ~ 0.2–0.1 s$^{-1}$ and $L_{Li}$ ~ 2–7×10$^7$ s$^{-1}$. For $\delta_{Rb}$ = -12 – -24 MHz, typical values are: $\kappa_{Rb}$ ~ 0.3–0.6 s$^{-1}$ and $L_{Rb}$ ~ 1–5×10$^7$ s$^{-1}$. We assume that these values measured from single species operation remain unchanged for two species operation.

To obtain the values of $\beta_{Li,Rb}$ and $\beta_{Rb,Li}$, both MOTs are allowed to load simultaneously. When a steady state is reached, $(dN_A/dt)$ of Eqs. (2) and (3) can be set to zero leading to:

$$\beta_{Li,Rb} = (L_{Li} - \kappa_{Li} \bar{N}_{Li}^\infty)/(\bar{n}_{Rb} \bar{N}_{Li}^\infty)$$

$$\beta_{Rb,Li} = (L_{Rb} - \kappa_{Rb} \bar{N}_{Rb}^\infty)/(\bar{n}_{Li} \bar{N}_{Rb}^\infty)$$

where, the "¯" is used to denote the steady-state number of atoms or density of MOTs when both species are simultaneously present.

The loss rate coefficients generally depend on the MOT parameters such as laser intensities and detuning [6, 7, 9]. We can use these dependences to understand the nature of the inelastic collisions that lead to trap loss, as has been done previously with losses for other species [6-19]. Several possible mechanisms have been identified, including radiative escape (RE), fine-structure changing collisions, hyperfine changing collisions, and molecule formation. In RE, atoms $A$ (in an excited electronic state, designated $A^*$) and $B$ (in its ground state) approach one another along an attractive potential energy curve. As their potential energy decreases, their kinetic energy (and velocity) increases. Spontaneous emission during the collision will then generate a scattered photon at a lower energy than that of the photon originally absorbed by $A$, with the difference in energy found as kinetic energy of the ground state atoms $A$ and $B$. If this energy is greater than the trapping potential for either $A$ or $B$, then one or both of these atoms can escape from the trap, contributing to the trap losses. In the Li-Rb system, Rb*-Li collisions (where the asterisk indicates the Rb is in the 5$p$

$^2P_{3/2}$ state) can result in RE, but the potential curves for Li*-Rb collisions (in which Li* designates Li in the $2p\ ^2P_{3/2}$ state), are repulsive, and RE is precluded. The spontaneous emission event can also leave atom *A* or *B* in the untrapped hyperfine ground state, also leading to trap loss, depending on the recovery rate of atoms in this state by the repump laser and the MOT trap depth. In the present work, the repump beams are relatively intense (leading to rapid recovery of these atoms) and the MOT trap depths are relatively high; hence, we expect that losses due to these processes are not significant. Collisions can also cause transitions between fine-structure states (of the excited states, since the ground state of an alkali metal atom has no fine structure) or hyperfine states (of the ground or excited state, although trap loss is more likely when changing hyperfine states in the ground state system, due to the larger hyperfine energy in the ground state than in the excited state). The energy difference between the fine-structure states or hyperfine states is transferred to kinetic energy of *A* and *B*, which can result in their loss from the system. However, as discussed above, we do not expect hyperfine changing collisions to be important in our system since the MOT trap depths (>1 K) are significantly higher than the energy associated with the hyperfine structure (maximum hyperfine energy is 3.04 GHz, i.e. 0.14 K, corresponding to the ground state hyperfine splitting of $^{85}$Rb). It is difficult to differentiate between fine structure changing collisions and RE on the basis of our measurements, and they are together referred to as losses due to Li-Rb* collisions. Finally, formation of a molecule *AB* by the colliding atoms results in loss of both species from the traps.

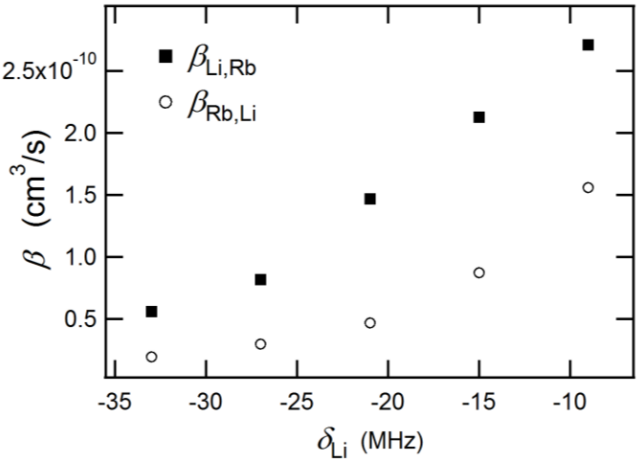

FIG. 6. The dependence of the loss rate coefficients on the detuning ($\delta_{Li}$) of the Li cooling laser. The Li repump detuning is fixed at -18 MHz, the Rb cooling laser detuning is fixed at -24 MHz and the Rb repump is resonant. The filled squares and open circles are the values of $\beta_{Li,Rb}$ and $\beta_{Rb,Li}$ respectively.

In Fig. 6, we show the dependence of $\beta_{Li,Rb}$ on the detuning $\delta_{Li}$ of the Li cooling beam from the $F = 2 \rightarrow F' = 3$ transition, with the detuning ($\delta_{Rb}$) of the Rb cooling beam held fixed at $\delta_{Rb}$ = -24 MHz. It is seen that the value of $\beta_{Li,Rb}$, characterizing the Rb-induced Li loss, decreases from ~ $2.7\times10^{-10}$ cm$^3$/s at $\delta_{Li}$ = -9 MHz to ~ $5.6\times10^{-11}$ cm$^3$/s at $\delta_{Li}$ = -33 MHz.

The detuning $\delta_{Li}$ affects three primary MOT characteristics [51-53], which in turn can affect the collisional loss rates. With increasing detuning $|\delta_{Li}|$, the trap depth increases, the temperature of the Li MOT increases and the population in the excited $2p\ ^2P_{3/2}$ state decreases. (The increasing trap depth and increasing temperature of the Li MOT with increasing $|\delta_{Li}|$ is uncommon among trapped atomic species. In most traps, such as Rb, the trap depth and temperature *decrease* with increasing detuning of the trapping laser.) The dependence of $\beta_{Li,Rb}$ on $\delta_{Li}$ shown in Fig. 6 is consistent with the variation of trap depth, but counter to the variation in temperature. As we increase $|\delta_{Li}|$, the increasing trap depth makes it more difficult for Li atoms to escape the trap, as reflected in the decreased loss rate coefficient $\beta_{Li,Rb}$. Conversely, we expect that the increasing temperature of the Li MOT with increasing $|\delta_{Li}|$ would manifest itself as an *increasing* $\beta_{Li,Rb}$. (An increase in the temperature implies an increase in the (average) velocity $v_{Li}$ of Li atoms. But $v_{Li}$ is nearly equal to the relative velocity of the colliding Li and Rb atoms, since the average velocity $v_{Rb}$ of the Rb atoms is expected to be much less than $v_{Li}$. This is because the typical temperature of the Rb MOT (few hundred µK) is much lower than the typical temperature (few mK) of the Li MOT, and the Rb atomic mass $m_{Rb}$ is much greater than the Li atomic mass $m_{Li}$.) In inelastic collisions, energy and momentum conservation during collisions require that a fraction $m_{Rb}/(m_{Li} + m_{Rb}) \approx 92\%$ of any released energy be deposited in the lighter Li atom after a Li-Rb collision. The gain in kinetic energy of Li atoms is much greater than that of Rb atoms and it is thus much more likely for a Li atom to leave the trap (typical Li MOT trap depth ~ 1K [51]) than a Rb atom (typical Rb MOT trap depth ~ 10K [19,54]). Another possible factor, the population of the Li excited $2p\ ^2P_{3/2}$ state, can also be ruled out because the interaction between an excited ($2p\ ^2P_{3/2}$) Li atom and a ground state ($5s\ ^2S_{1/2}$) Rb atom is repulsive [55] preventing the Li and Rb atoms from getting close enough where inelastic loss-inducing collisions can occur. Our observation that $\beta_{Li,Rb}$ increases with decreasing detuning $|\delta_{Li}|$, therefore, leads us to conclude that the variation in the Li MOT trap depth is more important than that of the temperature of the Li atoms.

Figure 6 also shows the dependence of $\beta_{Rb,Li}$, characterizing Li-induced Rb losses, on the detuning $\delta_{Li}$ of the Li cooling beam, with the detuning of the Rb cooling beam held fixed at $\delta_{Rb}$ = -24 MHz. The trend is similar to that of $\beta_{Li,Rb}$, with $\beta_{Rb,Li}$ being a factor of ~3 lower than $\beta_{Li,Rb}$. The trend cannot be attributed to an increase in Li MOT trap depth with increasing detuning since the Li MOT trap depth cannot play a role in determining the Rb losses. The dependence can also not be attributed to the increase in Li MOT temperature with increasing detuning, since that would imply an increase in $\beta_{Rb,Li}$ with increasing detuning, contrary to the experimental observation. In addition, as mentioned above, the population of the excited ($2p$ $^2P_{3/2}$) state Li atoms should play no role in determining $\beta_{Rb,Li}$ or $\beta_{Li,Rb}$. We speculate that this could be indicative of molecule formation, to be discussed in the following paragraphs. We note that the trend is actually reverse of that observed for $\beta_{Cs,Li}$ in [9], where the increase in $\beta_{Cs,Li}$ with increasing $|\delta_{Li}|$ was attributed to the increase in the temperature of Li atoms in the MOT.

In order to study the dependence of $\beta_{Li,Rb}$ and $\beta_{Rb,Li}$ on the detuning $\delta_{Rb}$ of the Rb cooling beam, the above measurements were repeated at a lower detuning of $\delta_{Rb}$ = -18 MHz and $\delta_{Rb}$ = -12 MHz. Within our experimental uncertainty, the values of $\beta_{Li,Rb}$ and $\beta_{Rb,Li}$ were comparable for all three values of $\delta_{Rb}$. The dependence of the trap depth and temperature of the Rb MOT on $|\delta_{Rb}|$, both of which decrease with increasing $|\delta_{Rb}|$, is opposite that of the Li MOT; while population in the excited $5p$ $^2P_{3/2}$ state decreases with increasing $|\delta_{Rb}|$ [19, 56]. We expect that the temperature of the Rb MOT, however, has little affect on the loss rate coefficients because the relative velocity of collisions is determined solely by the temperature of the much hotter Li MOT, as already mentioned. The interaction between ground state Li atoms and Rb atoms in the excited $5p$ $^2P_{3/2}$ state (denoted by Rb*) is attractive in nature and can aid in bringing the Li and Rb atoms close enough for loss inducing collisions to occur. The population of Rb in the excited $5p$ $^2P_{3/2}$ state decreases with increasing detuning $|\delta_{Rb}|$, which should reduce Rb-induced Li losses. We are unable to observe this effect in our collision induced loss measurements but this could be due to the relatively small range over which $\delta_{Rb}$ is varied. However, our observations, detailed below, while using a dark MOT for Rb clearly indicate that collisions between ground state Li and excited Rb* atoms account for majority of the atom losses observed.

Table 1. Values of loss rate coefficients (in cm$^3$/s) measured for conventional bright MOTs. Parameters involving $^6$Li are from [10] while those involving $^7$Li are measured in our experiment. There are no reported measurements for $^7$Li - $^{87}$Rb. The dependence of loss rate coefficients on the isotopes could be attributable to the binding energy of the highest energy vibrational state within the electronic potential well, or possibly differences in the hyperfine structures of different isotopes [14, 19]. In the final row, we present the ratios $\beta_{Li,Rb}/\beta_{Rb,Li}$ for these different isotopic systems.

|  | $^7$Li - $^{85}$Rb | $^6$Li - $^{85}$Rb | $^6$Li - $^{87}$Rb |
|---|---|---|---|
| $\beta_{Li,Rb}$ | 5.6×10$^{-11}$ | 4×10$^{-10}$ | 2.5×10$^{-10}$ |
| $\beta_{Rb,Li}$ | 2.0×10$^{-11}$ | 5×10$^{-11}$ | 1.7×10$^{-11}$ |
| $\beta_{Li,Rb}/\beta_{Rb,Li}$ | 2.8 | 8 | 15 |

The values of $\beta_{Li,Rb}$ and $\beta_{Rb,Li}$ for collisions between $^7$Li and $^{85}$Rb are being reported here for the first time. It is interesting to compare these with the values of other isotopes of Li and Rb (Table 1). The loss rate coefficients for $^6$Li and $^{85}$Rb have been reported to be $\beta_{^6Li,^{85}Rb}$ = 4×10$^{-10}$ cm$^3$/s and $\beta_{^{85}Rb,^6Li}$ = 5×10$^{-11}$ cm$^3$/s [10]. These values were measured for $\delta_{Rb}$ = -11 MHz and $\delta_{Li}$ = -34 MHz. At similar detuning, we find the loss rate coefficients for $^7$Li and $^{85}$Rb to be $\beta_{^7Li,^{85}Rb}$ = 5.6×10$^{-11}$ cm$^3$/s and $\beta_{^{85}Rb,^7Li}$ = 2.0×10$^{-11}$ cm$^3$/s. Thus the values of $\beta_{Rb,Li}$ are similar for the two cases but the value of $\beta_{Li,Rb}$ is significantly different. In contrast to the observation in reference [10], where $\beta_{^6Li,^{85}Rb}$ is an order of magnitude higher than $\beta_{^{85}Rb,^6Li}$, we observe that the values of $\beta_{^7Li,^{85}Rb}$ and $\beta_{^{85}Rb,^7Li}$ differ by only a factor of ~3. (Direct comparison of loss rate coefficients, $\beta_{Li,Rb}$ or $\beta_{Rb,Li}$, with those determined for other isotopic systems by other research groups, can be difficult due to uncertainties in the absolute determination of atomic densities or numbers, as well as subtle differences in laser powers or detunings. We therefore compare ratios $\beta_{Li,Rb}/\beta_{Rb,Li}$ for the different isotopic species, which we expect to be more reliable.) We speculate that the similarity in the values of $\beta_{^7Li,^{85}Rb}$ and $\beta_{^{85}Rb,^7Li}$ can be explained by the formation of LiRb molecules in the electronic ground state. LiRb molecules in the electronic ground state can be formed in the two-species MOT by spontaneous emission of excited state LiRb* molecules formed by collisions of Li and Rb* atoms. The molecules, being transparent to the MOT beams, cannot be trapped and both atoms are lost from the MOT. Since $\beta_{Li,Rb}$ and $\beta_{Rb,Li}$ are similar, but not identical ($\beta_{Li,Rb}/\beta_{Rb,Li} \sim 3$), this could indicate that about 30% of the collisions lead to molecule formation, resulting in the loss of a Li and Rb atom from the trap, while the majority of collisions lead to the loss of a Li atom alone. Subtle differences between the vibrational energies of excited potentials for different isotopic species could allow for differences in molecule formation rates. For example, isotopic changes in the vibrational energy spacing can change the binding energy of the highest bound vibrational level substantially, affecting the collision process. A difficulty with this explanation lies with dependence of this rate on $\delta_{Li}$ and $\delta_{Rb}$. The molecule formation rate should depend only on the Rb detuning (and not on Li detuning) because the Li-Rb* interaction is attractive while Li*-Rb is repulsive. In Fig. 6, only the Li* population is being changed, yet we see variation of both loss coefficients. If molecule formation is the only loss mechanism, then both $\beta_{Li,Rb}$ and $\beta_{Rb,Li}$ should have been individually constant as the Li detuning was changed. This is obviously not the case. The other way to think about it is that the molecules, if formed, are always lost from the trap, irrespective of the MOT trap depth/detuning.

In order to investigate whether the isotopic difference in the loss rate is due to subtle differences in molecule formation rates (in electronically excited states, denoted as LiRb*) induced by the Rb MOT light, we performed experiments to determine the least bound state of the

$^7$Li$^{85}$Rb* molecule. If the least bound state of the $^7$Li$^{85}$Rb* molecule lies very close to the dissociation asymptote (i.e. the D$_2$ line of Rb) then it is possible for the Rb MOT trapping light to induce $^7$Li$^{85}$Rb* molecule formation via photoassociation (PA). We performed traditional PA measurement using trap loss spectroscopy in order to determine the least bound state of the $^7$Li$^{85}$Rb* molecule [57]. We found that there is one electronic state, the 3(0$^+$) state, in the $^7$Li$^{85}$Rb* molecule for which the least bound vibrational level is within tens of MHz from the D$_2$ asymptote. The least bound state could not be observed directly by PA because the PA laser, being very close in frequency to the D$_2$ line, severely distorted the Rb MOT. Instead, we observed the more deeply bound vibrational levels of the 3(0$^+$) state and extrapolated, using the LeRoy-Bernstein [58] formula, to determine the least bound state. Such extrapolations are, in general, very accurate and we find the least bound level of the 3(0$^+$) state is bound by only ~12 (±10) MHz and the outer turning point for this level is at a very large internuclear separation of ~ 150 $a_0$ (1 $a_0$ ≈ 0.53 Å). We also note that the linewidth of the PA lines in the 3(0$^+$) state is found to be ~250 MHz, much larger than the uncertainty in the determination of the position of the least bound state. Since the least bound state is very close to the Rb (5$p$ $^2$P$_{3/2}$) dissociation asymptote and since ground state Li and Rb atoms can come closer than 150 $a_0$ (just for a reference, the $p$-wave centrifugal barrier is at ~100 $a_0$), it is highly possible that the Rb MOT trapping light leads to formation of electronically excited $^7$Li$^{85}$Rb* molecule by PA. Such $^7$Li$^{85}$Rb* molecules formed in the MOT could spontaneously decay to form $^7$Li$^{85}$Rb molecules in the electronic ground state (and hence be lost from the MOT). We tried detecting the ground state molecules by ionizing them using REMPI but no such molecules could be detected either in the MOT or with PA to any of the bound levels of the 3(0$^+$) state. One possible reason is that the $^7$Li$^{85}$Rb* molecules formed in the 3(0$^+$) state undergo predissociation leading to formation of free Li and Rb atoms with high kinetic energies both of which are then lost from the MOT. Irrespective of whether or not ground state $^7$Li$^{85}$Rb molecules are formed, this picture of formation of excited state $^7$Li$^{85}$Rb* molecules induced by the Rb trapping light partially explains why $\beta_{Li,Rb}$ and $\beta_{Rb,Li}$ are similar for collisions between $^7$Li and $^{85}$Rb. For other isotopic combinations (like $^6$Li and $^{85}$Rb/$^{87}$Rb in ref. [10]), the least bound state may not be so close to the Rb (5$p$ $^2$P$_{3/2}$) dissociation asymptote and formation of LiRb* molecule might be unlikely.

A few words about possible sources of errors in the measurement of $\beta_{Li,Rb}$ and $\beta_{Rb,Li}$ are warranted here. Random errors in the values of $\beta_{Li,Rb}$ and $\beta_{Rb,Li}$ are minimal and the trends seen in Fig. 6 are reproducible. The primary sources of error are the systematic errors arising from the uncertainties in the measurement of the number of atoms in and the sizes of the MOTs. The calculation of atom number requires the knowledge of the photon scattering rate which in turn depends on the intensity, polarization and detuning of the MOT cooling beams. The detuning is quite well determined in our experiments as is the intensity, but the polarization may not be perfect, and it varies through the MOT region due to interference effects between the six trapping beams. Other errors arise from the uncertainty in the solid angle subtended by the MOT at the collection lens and on the photon collection efficiency of the imaging system. Collectively these lead to an estimated systematic error of ~ 25% in the measurement of number of atoms trapped in the MOTs. The sizes (diameters) of the MOTs are estimated to be accurate within 15%. Together, these lead to an uncertainty of ~ 50% in the determination of $\beta_{^7Li,^{85}Rb}$ and $\beta_{^{85}Rb,^7Li}$. However, as noted earlier, these are systematic errors appearing in every measurement and hence the trends seen in Fig. 6 should not change significantly. Such uncertainties are typical in the measurement of loss rates [9].

### 3.2 Reduction of collision-induced losses in a dark MOT

As discussed earlier, the population of Rb atoms in the excited 5$p$ $^2$P$_{3/2}$ state can cause the loss of Li atoms from the MOT. The importance of the role of Rb*-Li collisions is further supported by our measurements of interspecies collision-induced losses with a dark MOT, also known as the dark spontaneous-force optical trap [59, 60]. In a dark MOT for Rb, the population in the excited 5$p$ $^2$P$_{3/2}$ state is reduced, and the trapped atoms primarily occupy the 5$s$ $^2$S$_{1/2}$ $F$=2 state. We obtain a dark Rb MOT by blocking the center of the Rb repumping beam with an opaque circular disc 6 mm in diameter. In addition, we detune the Rb repumping beam by +12 MHz from the $F = 2 \rightarrow F' = 3$ transition. We found that the use of an additional depumping beam, tuned to the $F = 3 \rightarrow F' = 2$ transition, was not required to reduce the Rb-induced Li losses.

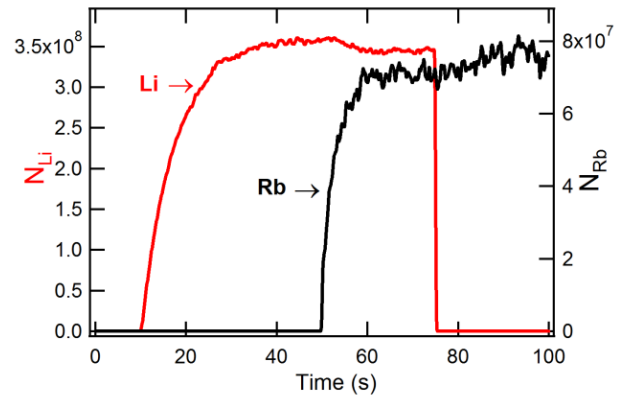

FIG. 7. (Color online) Simultaneous loading of a Li MOT (red, left axis) and dark Rb MOT (black, right axis). The Li MOT starts loading at $t = 10$ s and is allowed to reach steady state. At $t = 50$ s, the Rb dark MOT starts loading and a very small reduction in the Li MOT atom number is seen. At $t = 75$ s the Li MOT is blocked and a very small increase in the dark Rb MOT atom number is seen. The interspecies collision induced losses are greatly reduced when a dark Rb MOT is used.

Figure 7 shows the MOT loading curves when a Li MOT and a dark Rb MOT are simultaneously loaded where the losses are substantially reduced (also see Fig. 8). It is seen that the number of trapped atoms of one species is affected only slightly by the presence of the other species, thus preserving the densities also. The result clearly indicates that collisions of Li atoms with Rb atoms in the excited $5p\ ^2P_{3/2}$ state lead to the severe loss of Li atoms from the Li MOT, as speculated earlier.

### 3.3 Alternate method for the measurements of $\beta_{Li,Rb}$

The Li loss rate coefficient $\beta_{Li,Rb}$ can be measured using a different method. The Li MOT is first allowed to load and reach a steady state ($N_{Li,Max}$) in the absence of the Rb MOT. The Li atomic beam is then suddenly blocked (at $t = 0$) which suddenly changes the loading rate $L_{Li}$ to zero. The decrease in the Li MOT atom number ($N_{Li}$) can be described by Eq. (4) with $L_{Li}$ set to zero, and, the solution to the equation is $N_{Li} = N_{Li,Max} e^{-\kappa_{Li} t}$. In Fig. 8 we plot $\ln(N_{Li}/N_{Li,Max})$ vs. $t$ and extract $\kappa_{Li}$ (~ 1/7.5 s$^{-1}$) from the slope of the red curve.

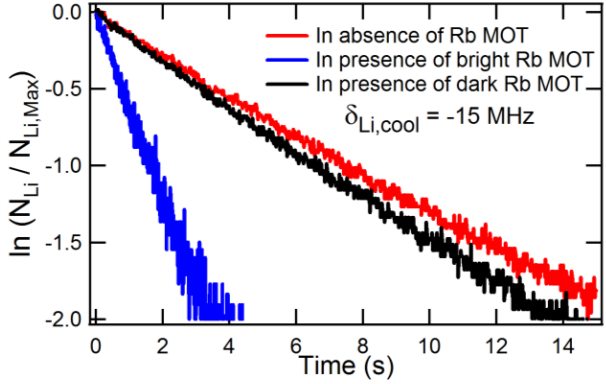

FIG. 8. (Color online) The decay of Li MOT atom number when the Li atomic beam is suddenly blocked at $t = 0$ s. The ratio $N_{Li}/N_{Li,Max}$ in log scale as a function of time after the atomic beam is blocked. In absence of the Rb MOT, the Li atom number decreases slowly with a lifetime ~ 7.5 s. In the presence of the bright Rb MOT, the Li atom number decreases much more quickly with a lifetime ~ 1.7 s. In the presence of the dark Rb MOT, the Li atom number decreases slowly with a lifetime ~ 6.5 s. This suppression of collision induced losses with a dark Rb MOT indicates that the Li losses are mainly due to collision of ground state Li atoms with excited state Rb* atoms.

Next, the Li MOT is allowed load in presence of the Rb MOT and it reaches a new steady state. The Li atomic beam is then blocked and the evolution of $N_{Li}$ can be described by Eq. (2) with $L_{Li}$ set to zero. The solution to the equation is $N_{Li} = N_{Li,Max} e^{-\kappa_{Li,Rb} t}$, where $\kappa_{Li,Rb} = \kappa_{Li} + \beta_{Li,Rb} n_{Rb}$. In Fig. 8 we plot $\ln(N_{Li}/N_{Li,Max})$ vs. $t$ and extract $\kappa_{Li,Rb}$ (~ 1/1.7 s$^{-1}$) from the slope of the blue curve. We then calculate $\beta_{Li,Rb}$ from the relation: $\beta_{Li,Rb} = (\kappa_{Li,Rb} - \kappa_{Li})/n_{Rb}$ (~ 1.5×10$^{-10}$ cm$^3$/s). The value $\beta_{Li,Rb}$ obtained using this method is consistent with the value obtained using the previous method. The value of $\beta_{Rb,Li}$ cannot be obtained using this alternate method since we do not have a shutter to instantaneously block the Rb atoms effusing out of the Rb dispenser. Note that this alternate method is relatively immune to small errors in the measurement of Li atom number and density. For completeness, we also plot the evolution of $N_{Li}$ in the presence of a dark Rb MOT and find that using a dark Rb MOT significantly reduces the Li losses and hence $\beta_{Li,Rb}$.

## 4. CONCLUSION

In this article, we report the measurement of collision-induced loss rate coefficients $\beta_{^7Li,^{85}Rb}$ and $\beta_{^{85}Rb,^7Li}$, and also discuss means to significantly suppress such collision induced losses. We first describe our dual-species magneto-optical trap (MOT) for simultaneous cooling and trapping of $^7$Li and $^{85}$Rb that allows us to simultaneously trap $\geq 5\times10^8$ $^7$Li atoms and $\geq 2\times10^8$ $^{85}$Rb atoms. We observe strong interspecies collision-induced losses in the MOTs which dramatically reduce the maximum atom number achievable in the MOTs. We measure the trap loss rate coefficients $\beta_{^7Li,^{85}Rb}$ and $\beta_{^{85}Rb,^7Li}$, and, from a study of their dependence on the MOT parameters, determine the major cause for such losses to be the Rb*-Li collisions. Our results provide valuable insights into ultracold collisions between $^7$Li and $^{85}$Rb, guide our efforts to suppress collision induced losses, and also pave the way for the production of ultracold $^7$Li$^{85}$Rb molecules [57,61].

### ACKNOWLEDGMENTS

Support during early stages of this work by the National Science Foundation (CCF-0829918), and through an equipment grant from the ARO (W911NF-10-1-0243) are gratefully acknowledged.

* sourav.dutta.mr@gmail.com
‡ elliottd@purdue.edu
† yongchen@purdue.edu